# Vibrational and electronic ultrafast relaxation of the nitrogen-vacancy centers in diamond

V. M. Huxter[1], T. A. A. Oliver[1], D. Budker[2], and G. R. Fleming[1]
[1]Department of Chemistry, University of California, Berkeley and Physical Bioscience Division, Lawrence Berkeley National Laboratory, Berkeley, CA, USA 94720
[2]Department of Physics, University of California, Berkeley and Nuclear Science Division, Lawrence Berkeley National Laboratory, Berkeley, CA, USA 94720

**Abstract.** Two dimensional electronic spectroscopy and transient grating measurements were performed, for the first time, on nitrogen-vacancy centers in diamond. These measurements reveal energy transfer and vibrational pathways with consequences for spin coherence.

## 1 Introduction

Nitrogen-vacancy (NV) centers in diamond, which consist of a nitrogen substitution with a nearest neighbor vacancy point defect in the carbon lattice, display spin coherences that last for up to one second at room temperature [1]. As a quantum solid-state system, NV centers are particularly promising due to their optically and magnetically addressable long spin coherences, fast spin manipulation [2] and coupling to adjacent electronic and nuclear spins [3]. These systems have wide ranging applications including solid-state qubits, ultrasensitive magnetometers, and ultra-resolution imaging.

In a diamond system containing NV centers (NV-diamond), the spin states can be initialized and read out by accessing an optical transition between triplet states due to a spin dependent intersystem-crossing pathway. While the transitions associated with the electronic state have been investigated in earlier work [4, 5], their ultrafast dynamics are completely unknown. In addition to the electronic levels, NV-diamond systems have significant phonon sidebands. These vibrational modes influence the decoherence of the spin state and the flow of energy through the system. Despite their importance to our understanding of NV-diamond, little is known about them beyond steady-state or low-frequency-modulation measurements. In order to elucidate the relaxation pathways of this solid-state quantum system, we present the first ultrafast optical measurements on NV-diamond. Using two dimensional electronic spectroscopy (2DES) and frequency resolved transient grating measurements (FRTG), the dynamics of both the electronic and vibrational relaxation pathways in NV-diamond are measured on ultrafast timescales, allowing for the creation of an energy-function map of the system.

## 2 Experimental

2DES measurements were performed using an apparatus described in detail previously [6]. A home-built Ti:sapphire oscillator/regenerative amplifier was used to generate pulses in the visible wavelength range using a non-collinear optical parametric amplifier. A diffractive optic was used to generate four passively phase stabilized beams that were focused into the sample using a box geometry. The signal radiated in the $k_s=-k_1+k_2+k_3$ phase-matching direction was spectrally dispersed and heterodyne-detected on a CCD. To obtain the 2DES data, the time delay between the first two interactions (the coherence time, $\tau$, also designated as time delay *1*) was scanned and the collected interferograms were Fourier transformed to produce the frequency-domain 2D spectrum. Population dynamics were obtained by repeatedly scanning $\tau$ for different waiting times, $T$ (the time delay between the second and third interactions, also designated as time delay *2*). FRTG measurements were performed using the same experimental apparatus by setting the coherence time to zero and scanning the waiting time. The FRTG experiment was performed to provide a complementary measure of the population dynamics with finer time resolution than the equivalent result obtained from the 2DES measurement.

The NV-diamond sample was prepared using a high pressure-high temperature method with an approximately 50 ppm nitrogen concentration. The sample was irradiated with 3 MeV electrons for a radiation dose of $2 \times 10^{19}$ cm$^{-2}$ and annealed at 1325 K for two hours. This produced negatively charged NV-diamond centers with an optical density at room temperature of 0.06 at the zero-phonon line (638 nm, 15700 cm$^{-1}$) with an optical pathlength of 20 μm, as shown in Figure 1(c). The 2DES and FRTG experiments were performed using 60 to 80 nm full width half maximum spectral bandwidth pulses with central wavelengths of 590, 600, 625 and 640 nm (only the 625 nm data is shown here) and compressed to near transform limited sub 20 fs pulses at the sample position.

## 3 Results and Discussion

A schematic of the NV-diamond energy level structure is shown in Figure 1(a). The $^3A$ and $^3E$ states are split into $m_s = \pm 1$ and $m_s = 0$ spin sublevels. While the spin states are not directly observable in the electronic optical measurements, they are affected by the electronic and vibrational transitions that are the subject of the current work. Singlet states with an energy gap corresponding to the near infrared are also shown in the energy-level diagram but they do not play a role in the current measurement. The absorption spectrum of the NV-diamond sample used in the experiment is shown in Figure 1(c). The zero-phonon line (ZPL), which corresponds to transitions between the ground and excited state occurring without the assistance of a vibration, is at 638 nm and persists at room temperature in NV-diamond. The broad feature to the high-energy side of the ZPL is the phonon sideband that spans more than 300 meV or more than 10 times $kT$ at room temperature. In the linear absorption spectrum, some vibronic structure is apparent in the phonon sideband. This structure is clearly shown in Figure 1(b), which is the second derivative of the absorption spectrum.

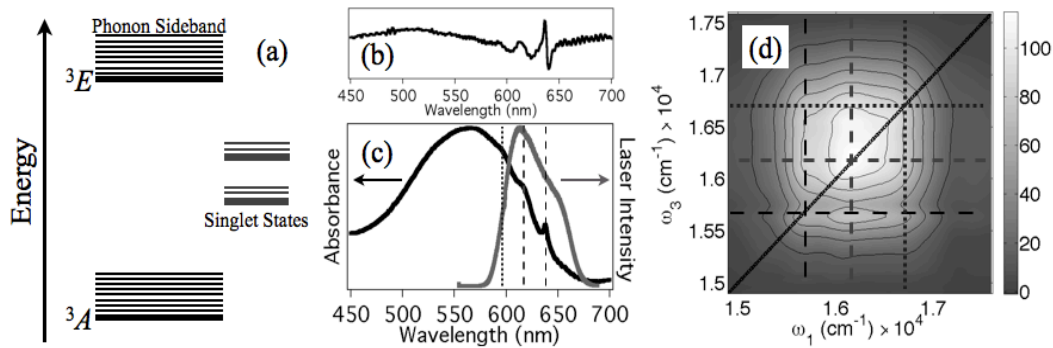

Figure 1 (a) Energy level schematic of the negatively charged NV center in diamond. (b) Second derivative of the absorption spectrum, emphasizing the vibronic structure of the phonon sideband. (c) Absorption spectrum of the NV-diamond sample and the laser spectrum centered at 625 nm used to produce the 2DES data shown in (d). Panel (d) is an absolute value 2DES plot of NV-diamond with a waiting time of 5 ps. The 2DES signal in (d) plots the detection axis, $\omega_3$, versus the absorption axis, $\omega_1$. The lines on panels (c) and (d) indicate the ZPL and the first and second vibronic peak. See text for additional details.

Figure 1(d) is the absolute value total 2DES measurement obtained with a laser spectrum centered at 625 nm [plotted with the NV-diamond absorption spectrum in Figure 1(c)] for a waiting time of 5 ps. At shorter waiting times, the energy moves through the system alternating between the peaks as expected for vibrational transitions. Several distinct vibrational modes were observed in the data collected with the laser spectrum centred at 625 nm. These same vibrational modes also appeared in the FRTG signal. The observed vibrational spectrum of the NV-diamond sample is in excellent agreement with theoretical predictions [7-9] and is dominated by local vibrational modes associated directly with the NV defect.

After a waiting time of approximately 8 ps, the major directly observable vibrational modes have been damped out. However, clear cross peaks at energies associated with the first and second vibronic transitions persisted for waiting times of tens of picoseconds, indicating that relaxation from the phonon side band and emission via the ZPL is shown to occur primarily through the vibronic levels. The absolute-value spectrum, shown in Figure 1(d), revealed the diagonal contribution of the ZPL transition along with cross peaks associated with energy transfer from the two lowest-energy vibronic modes. Even for the longest waiting times measured in this experiment (100 ps), emission from the phonon side band through the ZPL was primarily mediated by the vibrational modes.

## 4 Conclusion

The first ultrafast nonlinear measurements were performed on NV-diamond samples, revealing the dynamics of energy transfer and vibrational relaxation. Vibrational modes were directly observed in the 2DES and the FRTG measurements for waiting times of less than 8 ps. Several distinct vibrations associated with the NV centers were observed and found to be consistent with strong coupling to local vibrational modes, including those associated with symmetry lowering lattice distortions. The vibronic cross peaks in the data presented in Figure 1(d), showed direct evidence for vibrationally assisted transfer from the phonon sideband and emission though the ZPL, which persisted for at least 100 ps.

## 5 Acknowledgements

The authors thank Adam Gali for suggesting ultrafast measurements with NV-diamond and Andrey Jarmola for preparing the NV-diamond sample. V.M.H. thanks the National Science and Engineering Research Council of Canada for a postdoctoral fellowship. D.B. was supported by NSF, IMOD, and the AFOSR/DARPA QuASAR program. The work by V.M.H., T.A.A.O. and G.R.F. was supported by NSF grant CHE-1012168.